\documentclass[letterpaper,12pt]{article}

\usepackage[english]{babel}
\usepackage{theorem}
\theoremheaderfont{\itshape\bfseries}
{\theorembodyfont{\itshape}
	
	\newtheorem{lemma}{\textbf{Lemma}}
	\newtheorem{definition}{\textbf{Definition}}
	\newtheorem{theorem}{\textbf{Theorem}}
	
	\newtheorem{remark}{\textbf{Remark}}

}

\usepackage{color}
\usepackage{newtxtext,newtxmath}

\newcommand{\T}{^{\mbox{\tiny T}}}
\usepackage{url}

\DeclareMathOperator{\diag}{diag}

\DeclareMathOperator{\Ker}{ker}
\DeclareMathOperator{\Span}{span}
\newcommand{\R}{\mathbb{R}}

\let\geq\geqslant

\newenvironment{proof}[1][Proof]%
{\par\noindent\textit{#1:\ }}%
{\hspace*{\fill} \rule{6pt}{6pt}}
\newenvironment{proof*}[1][Proof]%
{\par\noindent\textit{#1:\ }}{}

\newenvironment{system}[1]%
{\setlength{\arraycolsep}{0.5mm}\left\{ \;
	\begin{array}{#1}}%
	{\end{array} \right.}
\newenvironment{system*}[1]%
{\setlength{\arraycolsep}{0.5mm}
	\begin{array}{#1}}%
	{\end{array}}

\usepackage{tikz}
\usetikzlibrary{shapes,calc,arrows,patterns,decorations.pathmorphing
	,decorations.markings}
\usetikzlibrary{arrows.meta}

\begin{document}
	
\title{Weak synchronization in heterogeneous multi-agent systems}
	
\author{Anton A. Stoorvogel, Ali Saberi, and Zhenwei Liu
	\thanks{Anton A. Stoorvogel is with Department of Electrical
		Engineering, Mathematics and Computer Science, University of
		Twente, P.O. Box 217, Enschede, The Netherlands (e-mail:
		A.A.Stoorvogel@utwente.nl)}
	\thanks{Ali Saberi is with
		School of Electrical Engineering and Computer Science, Washington
		State University, Pullman, WA 99164, USA (e-mail: saberi@wsu.edu)}
	\thanks{Zhenwei Liu is with College of Information Science and
		Engineering, Northeastern University, Shenyang 110819,
		China (e-mail: liuzhenwei@ise.neu.edu.cn)} 
} 

\maketitle

\begin{abstract}	
  In this paper, we propose a new framework for
synchronization of heterogeneous multi agent system which we refer
to as weak synchronization. This new framework of synchronization is
based on achieving the network stability in the absence of any
information on communication network including the
connectivity. Here by network stability, we mean that in the basic
setup of a multi-agent system, we require that the signals exchanged
over the network converge to zero. As such if the network happens to
have a directed spanning tree then we obtain classical
synchronization. Moreover, we design protocols which achieve weak
synchronization for any network without making any kind of
assumptions on communication network. If the network happens to have
a directed spanning tree, then we obtain classical
synchronization. However, if this is not the case then we describe
in detail in this paper what kind of synchronization properties are
preserved in the system and the output of the different agents can
behave.
\end{abstract}

\section{Introduction}

Multi-agent systems have been extensively studied over the past 20
years. Initiated by early work such as
\cite{ren-atkins,saber-murray2}, although the roots can be found in
much earlier work \cite{wu-chua2}, it has become an active research
area. But the realization that control systems often consist of many
components with limited or restricted communication between them was
already studied in the area of decentralized control, see
e.g. \cite{siljak,corfmat-morse2}. Applications are for instance
systems with many generators connected through a grid or traffic
applications such as platoons of cars. The fallacy of early
decentralized control is that it often created a specific agent which has
a kind of supervisory role while other agents ensure communication to
and from this supervisory agent. This approach turned out to be highly
sensitive to failures in the network.  Multi-agent systems created a different
type of structure in these networks where all agents basically have a
similar role towards achieving synchronization in the
network. However, early work still heavily relied on knowledge of the
network.

Later it was established that the protocols designed for a multi-agent
systems would work for any network structure satisfying some underlying
assumptions such as lower or upper bounds on the spectrum of the Laplacian matrix
associated to the graph describing the network structure. This
suggested some form of robustness against changes in the network.
However, this idea still has two major flaws:
\begin{itemize}
	\item Firstly, if the network is unknown, then we can never check
	whether these assumptions are actually satisfied and hence we do not
	know whether we will achieve synchronization.
	\item Secondly, changes in the network can have significant effect on
	the bounds that have been used. It is easily seen that a few links
	failing might yields a network that fails connectivity
	properties. But also \cite{tegling2018fundamental} showed that these
	lower bounds on the eigenvalues of the Laplacian almost always
	converge to zero when the network gets large makes these assumptions
	impossible to guarantee.
\end{itemize}
In recent years scale-free protocols have been studied, see for
instance \cite{liu-nojavanzedah-saberi-2022-book} and references
therein. These protocols get rid of all assumptions on the network
such as these bounds on the eigenvalues of the Laplacian. However, it
still requires that the network is strongly connected or has a direct
spanning tree. This actually still inherently has some of the
difficulties presented before. How can we check if this connectivity
is present in the network? Secondly, what happens in case of a fault
in the network that makes the network fail this assumption.

In the basic setup of a multi-agent system, the signals exchanged over
the network converge to zero whenever the network synchronizes. So the
fact that the network communication dies out over time is a weaker
condition than output synchronization. We will refer to this weaker
condition as weak synchronization in this paper. We will consider
heterogeneous agents in this paper but the concept equally applies to
homogeneous networks.

It turns out that if we have a linear scale-free protocol then
synchronization implies weak synchronization. But, more importantly,
if the network has a directed spanning tree then the converse
implication is true: weak synchronization implies classical
synchronization.

We can therefore design protocols which achieve weak synchronization
for any network without making any kind of assumptions. If the network
happens to have a directed spanning tree then we obtain classical
synchronization. However, if this is not the case then we describe in
detail in this paper what kind of synchronization properties are
preserved in the system. For applications this kind of weak synchronization what
one would hope for. If the cars in a platoon lose connectivity between
two subgroups because their distance has become too large the protocols
will still achieve synchronization in both of these groups. If in a power
system the connectivity between two subgroups is lost, each of these groups
will internally achieve synchronization but, obviously, no global
synchronization will be achieved.

\section{Communication network and graph}

To describe the information flow among the agents we associate
a {weighted graph} $\mathcal{G}$ to the communication network. The
weighted graph $\mathcal{G}$ is defined by a triple $(\mathcal{V},
\mathcal{E}, \mathcal{A})$ where $\mathcal{V}=\{1,\ldots, N\}$ is a
node set, $\mathcal{E}$ is a set of pairs of nodes indicating
connections among nodes, and $\mathcal{A}=[a_{ij}]\in
\mathbb{R}^{N\times N}$ is the weighted adjacency matrix with non
negative elements $a_{ij}$. Each pair in $\mathcal{E}$ is called an
{edge}, where $a_{ij}>0$ denotes an edge $(j,i)\in \mathcal{E}$ from
node $j$ to node $i$ with weight $a_{ij}$. Moreover, $a_{ij}=0$ if
there is no edge from node $j$ to node $i$. We assume there are no
self-loops, i.e.\ we have $a_{ii}=0$. A {path} from node $i_1$ to
$i_k$ is a sequence of nodes $\{i_1,\ldots, i_k\}$ such that $(i_j,
i_{j+1})\in \mathcal{E}$ for $j=1,\ldots, k-1$. A {directed tree} is a
subgraph (subset of nodes and edges) in which every node has exactly
one parent node except for one node, called the {root}, which has no
parent node. A {directed spanning tree} is a subgraph which is a
directed tree containing all the nodes of the original graph. If a
directed spanning tree exists, the root of this spanning tree has a
directed path to every other node in the network \cite{royle-godsil}.

For a weighted graph $\mathcal{G}$, the matrix $L=[\ell_{ij}]$
with
\[ \ell_{ij}=
\begin{system}{cl} \sum_{k=1}^{N} a_{ik}, & i=j,\\ -a_{ij}, &
	i\neq j,
\end{system}
\]
is called the {Laplacian matrix} associated with the graph
$\mathcal{G}$. The Laplacian matrix $L$ has all its eigenvalues in the
closed right half plane and at least one eigenvalue at zero associated
with right eigenvector $\textbf{1}$ \cite{royle-godsil}. The zero
eigenvalues of Laplacian matrix is always semi simple, i.e. its
algebraic and geometric multiplicities coincides. Moreover, if the
graph contains a directed spanning tree, the Laplacian matrix $L$ has
a single eigenvalue at the origin and all other eigenvalues are
located in the open right-half complex plane \cite{ren-book}.

\begin{figure}[ht] 
	\includegraphics[width=8cm]{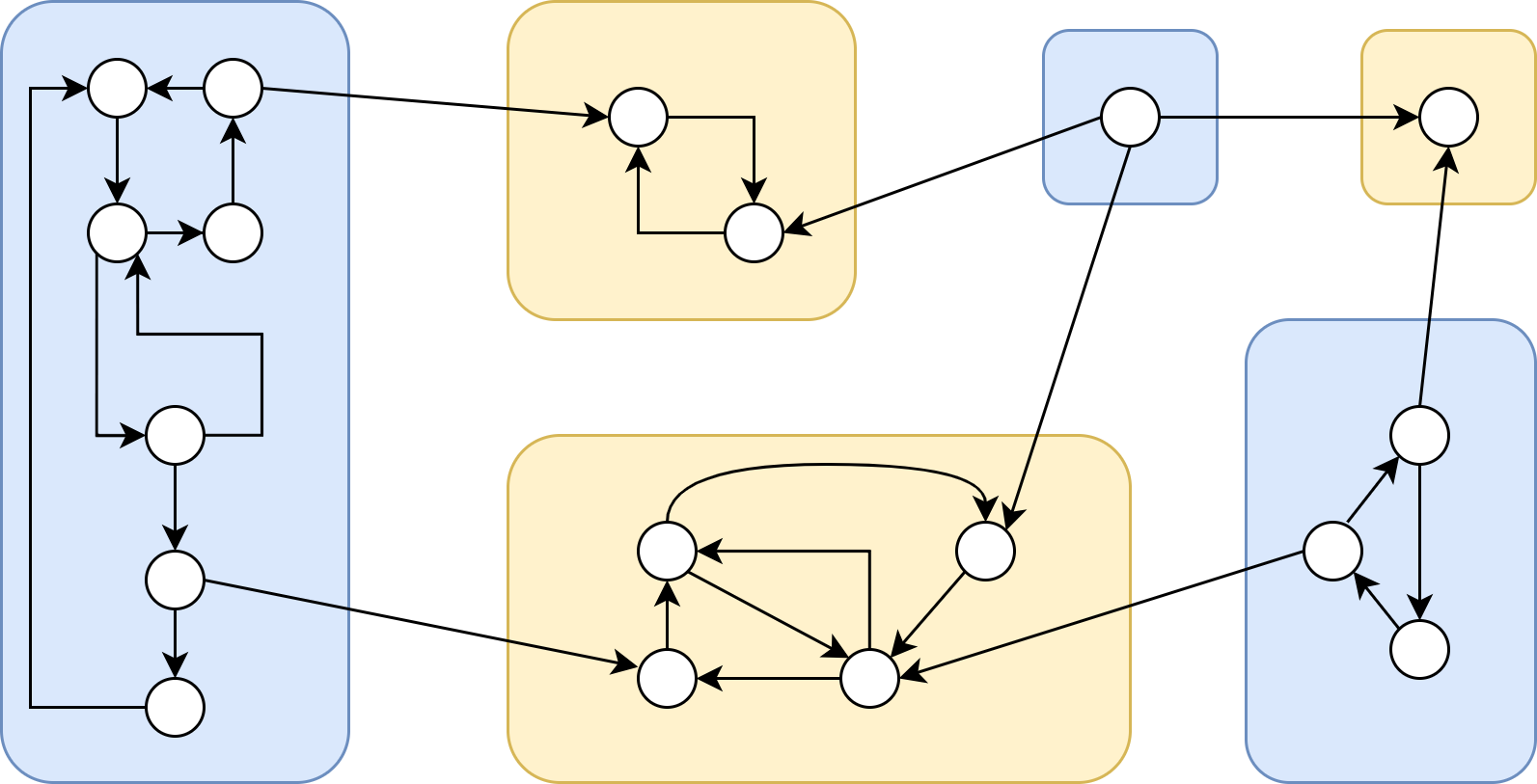}
	\centering
	\caption{A directed communication network and its bicomponents.}\label{f1}
\end{figure}

\begin{figure}[ht] 
	\includegraphics[width=8cm]{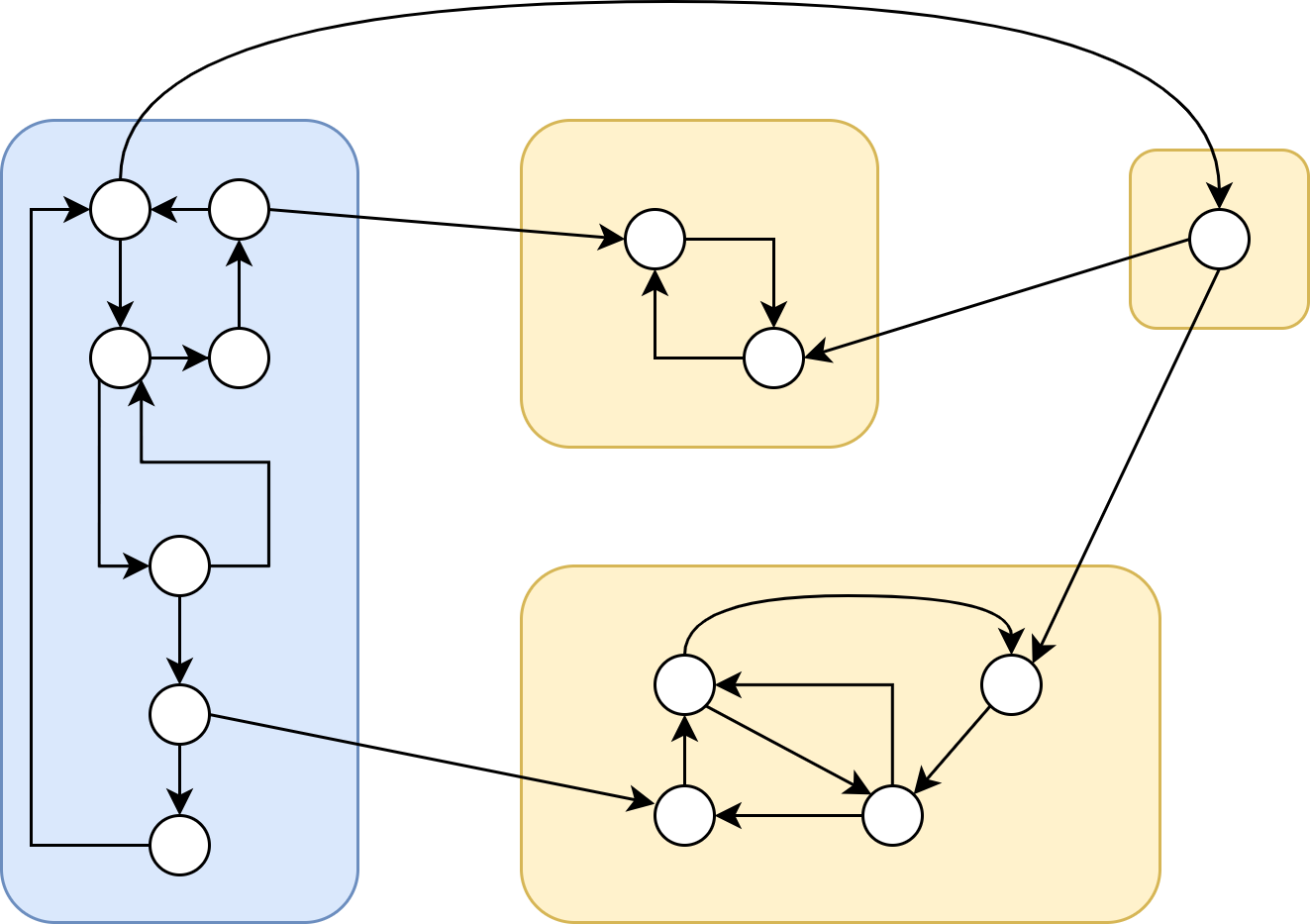}
	\centering
	\caption{A directed communication network with a spanning tree and its bicomponents.}\label{f2}
\end{figure}

A directed communication network is said to be strongly connected if
it contains a directed path from every node to every other node in the
graph. For a given graph $\mathcal{G}$ every maximal (by inclusion)
strongly connected subgraph is called a bicomponent of the graph. A
bicomponent without any incoming edges is called a basic
bicomponent. Every graph has at least one basic bicomponent. A network
has one unique basic bicomponent if and only if the network contains
a directed spanning tree. In general, every node in a network can be
reached by at least one basis bicomponent, see \cite[page
7]{stanoev-smilkov-2013}. In Fig. \ref{f1} a directed communication
network with its bicomponents is shown. The network in this figure
contains 6 bicomponents, 3 basic bicomponents (the blue ones) and 3
non-basic bicomponents (the yellow ones). In Fig. \ref{f2} a directed communication
network with its bicomponents is shown. The network in this figure
contains 4 bicomponents but only one basic bicomponent (the blue one).

In the absence of a directed spanning tree, the Laplacian matrix of
the graph has an eigenvalue at the origin with a multiplicity $k$
larger than $1$. This implies that it is a $k$-reducible matrix and
the graph has $k$ basic bicomponents.  The book \cite[Definition
2.19]{wu-book} shows that, after a suitable permutation of the nodes,
a Laplacian matrix with $k$ basic bicomponents can be written in the
following form:
\begin{equation}\label{Lstruc}
	L=\begin{pmatrix} L_0 & L_{01}
		& \cdots & \cdots & L_{0k} \\ 0 & L_1 & 0 & \cdots & 0 \\ \vdots & \ddots &
		\ddots & \ddots & \vdots \\ \vdots & & \ddots & L_{k-1} & 0 \\ 0 &
		\cdots & \cdots & 0 & L_k
	\end{pmatrix}
\end{equation}
where $L_1,\ldots, L_k$ are the Laplacian
matrices associated to the $k$ basic bicomponents in our
network. These matrices have a simple eigenvalue in $0$ because they
are associated with a strongly connected component. On the other hand,
$L_0$ contains all non-basic bicomponents and is a grounded Laplacian
with all eigenvalues in the open right-half plane. After all, if $L_0$
would be singular then the network would have an additional basic
bicomponent.

\section{Weak synchronization of MAS}

In this section, we introduce the concept of weak synchronization for
heterogeneous MAS. Consider $N$ heterogeneous agents
\begin{equation}\label{system}
	\begin{system*}{cl}
		x_i^+ &= A_ix_i +B_i u_i,  \\
		y_i &= C_ix_i,
	\end{system*}
\end{equation}
where $x_i\in\mathbb{R}^{n_i}$, $u_i\in\mathbb{R}^{m_i}$ and
$y_i\in\mathbb{R}^{p}$ are the state, input, output of agent $i$th for
$i=1,\ldots, N$. In the aforementioned presentation, for
continuous-time systems, $x_i^+(t) = \dot{x}_i(t)$ with
$t \in \mathbb{R}$ while for discrete-time systems, $x_i^+(t) = x_i(t+1)$
with $t \in \mathbb{Z}$.

The communication network provides agent $i$ with the following
information which is a linear combination of its own output relative
to that of other agents
\begin{equation}\label{zeta1}
	\zeta_i=\sum_{j=1}^{N}a_{ij}(y_i-y_j),
\end{equation}
where $a_{ij}\geq 0$ and $a_{ii}=0$. The communication topology of the
network can be described by a weighted and directed graph
$\mathcal{G}$ with nodes corresponding to the agents in the network
and the weight of edges given by coefficient $a_{ij}$. In terms of the
coefficients of the associated Laplacian matrix $L$, $\zeta_i$ can be
rewritten as
\begin{equation}\label{zeta}
	\zeta_i= \sum_{j=1}^{N}\ell_{ij}y_j.
\end{equation}
We denote by $\mathbb{G}^N$ the set of all graphs with $N$ nodes.  We
also introduce a possible additional localized information exchange among agents and their neighbors, i.e.\
each agent $i\in\{1,\ldots,N\}$ has access to {the} localized
information, denoted by $\hat{\zeta}_i$, of the form
\begin{equation}\label{zeta2}
	\hat{\zeta}_i=\sum_{j=1}^{N}a_{ij}(\eta_i-\eta_j),
\end{equation}
where $\eta_i$ is a variable produced internally by the protocol of
agent $i$ and to be defined later. Finally we might have introspective
agents which implies that
\begin{equation}\label{mop}
	y_{i,m} = C_{i,m} x_i
\end{equation}
is available to the protocol. 

Our protocols are of the form:
\begin{equation}\label{protocol}
	\begin{system*}{ccl}
		\xi_i^+ &=& K_i\xi_i + L_i \zeta_i + L_{i,e} \hat{\zeta}_i +
		L_{i,m} y_{i,m}\\
		u_i &=& M_i \xi_i \\
		\eta_j &=& N_i \xi_i
	\end{system*}
\end{equation}
For an agent $i$ which is introspective we might have $L_{i,m}\neq 0$
while for non-introspective agents we have that $L_{i,m}= 0$. Similar,
for an agent $i$ where extra communication is available of the form
\eqref{zeta2} we might have $L_{i,e}\neq 0$ while for agents without
extra communication we have $L_{i,e}=0$. 

In the following, we introduce the concepts of network stability and weak synchronization
that is vastly different from output synchronization.

\begin{definition}[\textbf{Network stability}]
	Consider a multi-agent network described by \eqref{system}, \eqref{zeta1},
	\eqref{zeta2}, \eqref{mop}, and \eqref{protocol}. If this network satisfies 
	\[
	\zeta_i(t)=\sum_{j=1}^{N}a_{ij}(y_i-y_j)\to 0
	\]
	as $t\to \infty$, for any $i \in \{1,\ldots, N\}$ and for all
	possible initial conditions, then it can be called a stable network.  
\end{definition}
\begin{definition}\label{wea}
	Consider an MAS described by \eqref{system}, \eqref{zeta1},
	\eqref{zeta2}, \eqref{mop}, and protocols \eqref{protocol}.  We have:
	\begin{itemize}
		\item The multi-agent network achieves output synchronization if the outputs of
		the respective agents satisfy:
		\begin{equation}\label{syncod}
			y_i(t)-y_j(t) \rightarrow 0
		\end{equation}
		as $t\rightarrow \infty$ for any $i,j\in \{1,\ldots, N\}$ and for
		all possible initial conditions.
		\item The multi-agent network achieves \textbf{weak synchronization} if the network
		is stable, i.e., 
		\[
		\zeta_i(t)\to 0
		\]
		as $t\to \infty$, for any $i \in \{1,\ldots, N\}$ and for all
		possible initial conditions.
	\end{itemize}
\end{definition}

Next, we present two lemmas to explain the difference between these two
kind of synchronization

\begin{lemma}\label{xxxxx}
	Consider an MAS described by \eqref{system}, \eqref{zeta1},
	\eqref{zeta2}, \eqref{mop}, and protocols \eqref{protocol}. In that case output synchronization
	implies weak synchronization.
\end{lemma}

\begin{proof}
	If $y_i(t)-y_j(t)\rightarrow 0$ then we have:
	\[
	\sum_{j=1}^{N}a_{ij}(y_i(t)-y_j(t))=\zeta_i(t) \rightarrow 0
	\]
	as $t\rightarrow \infty$ for all $i=1,\ldots, N$ which immediately
	implies weak synchronization.
\end{proof}

\begin{lemma}\label{xxxxx2}
	Consider an MAS described by \eqref{system}, \eqref{zeta1}, \eqref{zeta2},
	and \eqref{mop}. Assume the protocols \eqref{protocol} achieve weak
	synchronization. In that case:
	\begin{itemize}
		\item If the network contains a directed spanning tree then we
		always achieve output synchronization.
		\item If the network does not contain a directed spanning tree then
		we only achieve output synchronization for all initial conditions
		in the trivial case where $y_i(t)\rightarrow 0$ for all
		$i=1,\ldots,N$.
	\end{itemize}
\end{lemma}

\begin{proof}
	If the graph has a directed spanning tree then the associated
	Laplacian matrix has the property that
	$\Ker L = \Span \{ \textbf{1}_N \}$. This implies there
	exists a matrix $V\in\R^{(N-1)\times N}$ such that
	\[
	VL = \begin{pmatrix} I & -\textbf{1} \end{pmatrix}
	\]
	since $\Ker L = \Ker \begin{pmatrix} I & -\textbf{1}_{N-1} \end{pmatrix}$
	
	If we achieve weak synchronization we have that
	\[
	\zeta(t) = (I\otimes L) y(t) \rightarrow 0
	\]
	as $t\rightarrow \infty$ where:
	\[
	\zeta=\begin{pmatrix} \zeta_1 \\ \vdots \\
		\zeta_N \end{pmatrix},\qquad
	y=\begin{pmatrix} y_1 \\ \vdots \\
		y_N \end{pmatrix},\qquad
	\] 
	This implies
	\[
	(I\otimes V) \zeta(t) = \begin{pmatrix} y_1(t)-y_N(t) \\ \vdots \\
		y_{N-1}-y_N(t)  \end{pmatrix} \rightarrow 0
	\]
	as $t\rightarrow \infty$ which clearly implies output
	synchronization is achieved.
	
	Next we consider the case that the network doesn't contain a
	directed spanning tree. In that case, we have at least two basic
	bicomponents. By construction the behavior within a basic
	bicomponent is not influenced by the behavior of the rest of the
	network. Moreover, within this basic bicomponent we have a strongly
	connected graph. Hence within a basic bicomponent weak
	synchronization implies output synchronization. Assume within a
	basic bicomponent consisting of nodes $k_1,\ldots, k_i$ we have, for
	certain initial conditions, output synchronization such that
	\[
	y_{k_a}(t) - y_{\text{ss}}(t) \rightarrow 0,\qquad y_{\text{ss}}(t)
	\nrightarrow 0
	\]
	as $t\rightarrow \infty$ for all $a \in 1,\ldots,i$. Take a node $j$ within another basic
	bicomponent. If we have output synchronization then this would
	imply:
	\[
	y_{k_a}(t) - y_j(t) \rightarrow 0
	\]
	as $t\rightarrow \infty$  which is equivalent to
	\begin{equation}\label{jass}
		y_j(t) - y_{\text{ss}}(t) \rightarrow 0
	\end{equation}
	as $t\rightarrow \infty$. But if we multiply all initial conditions in the first basic
	bicomponent by a factor $2$ we get:
	\begin{equation}\label{kass}
		y_{k_a}(t)-2y_{\text{ss}}(t) \rightarrow 0
	\end{equation}
	as $t\rightarrow \infty$. On the other hand, if we keep the
	initial conditions for the second basic bicomponent the same
	then we would still have \eqref{jass}.  Clearly, \eqref{jass}
	and \eqref{kass} contradict output synchronization given that
	$y_{\text{ss}}(t) \nrightarrow 0$ as $t\rightarrow
	\infty$. Therefore we obtain our result by contradiction.
\end{proof}

\section{Dynamic behavior of the output of the agents  in a MAS given weak synchronization}

In this section, we focus on the behavior of the output of the agents
of a MAS for given protocols which achieves weak synchronization. We
have following theorem.

\begin{theorem}\label{thm_dy}
	Consider a MAS with agent dynamics \eqref{system} with protocols of
	the form \eqref{protocol} which achieves weak synchronization as
	defined by Definition \ref{wea}.
	
	Assume the network does not have a directed spanning tree which
	implies that the graph has $k>1$ basic bicomponents. Then, weak
	synchronization implies:
	\begin{itemize}
		\item Within basic bicomponent $\mathcal{B}_i$ for some
		$i\in \{1,\ldots,k\}$ the output of the agents synchronize and
		converge to the trajectories $y^i_s$.
		\item An agent $j$ which is not part of any of the basic
		bicomponents synchronizes to a trajectory $y_{j,s}$,
		\begin{equation}\label{ys}
			y_{j,s}=\sum_{i=1}^k\, \beta_{j,i} y^i_s 
		\end{equation}
		where the coefficients $\beta_{j,i}$ are nonnegative, satisfy:
		\begin{equation}\label{betasum}
			1=\sum_{i=1}^k\, \beta_{j,i} 
		\end{equation}
		and only depend on the parameters of the network and do not
		depend on any of the initial conditions.
	\end{itemize}
\end{theorem}

\begin{proof}
	Assume the Laplacian of the network is of the form
	\eqref{Lstruc}. Denote by $M_1,\ldots M_k$ the number of agents in
	the $k$ basic bicomponents respectively while $M_0$ is the number of
	agents not contained in a basic bicomponent.
	
	Since the matrices $L_1,\ldots, L_k$ have a one-dimensional
	nullspace while $L_0$ is nonsingular, there exists
	$\{ \beta_1,\ldots \beta_k\} \in \R^{M_0}$ such that the nullspace
	of $L$ is given by the image of the following matrix:
	\begin{equation} \label{betamatrix}
		\begin{pmatrix}
			\beta_1           & \beta_2          & \cdots  & \beta_k \\
			\textbf{1}_{M_1}  & 0                & \cdots  & 0       \\
			0                 & \textbf{1}_{M_2} & \ddots  & \vdots  \\
			\vdots            & \ddots           & \ddots  & 0       \\
			0                 & \cdots           & 0       & \textbf{1}_{M_k}
		\end{pmatrix}
	\end{equation}
	We define scalars $\beta_{1,i},\ldots , \beta_{M_0,i}$ such that:
	\[
	\beta_i = \begin{pmatrix} \beta_{1,i} \\ \vdots \\
		\beta_{M_0,i} \end{pmatrix}.
	\]
	Since we know $\textbf{1}_N$ is an element of the nullspace of $L$ we
	find that:
	\[
	\sum_{i=1}^k \beta_i = \textbf{1}_{M_0}
	\]
	which yields \eqref{betasum}. Using that
	\[
	\beta_i = -L_0^{-1} L_{0i}\textbf{1}_{M_i},
	\]
	we find that all coefficients of $\beta_i$ are nonnegative. This
	follows since the structure of the Laplacian guarantees that all
	coefficients of $L_{0i}$ are nonpositive while all coefficients of
	$L_{0i}^{-1}$ are nonnegative. The latter follows from \cite[Theorem
	4.25] {qu-book-2009}.
	
	Assume $\tau^i_1,\ldots \tau^i_{M_i}$ are the agents contained in basic
	bicomponent $i$. Then
	\[
	(I\otimes L) \begin{pmatrix} y_1(t) \\ \vdots \\
		y_N(t) \end{pmatrix} \rightarrow 0
	\]
	(which follows from weak synchronization) implies:
	\[
	(I\otimes L_i) \begin{pmatrix} y_{\tau^i_1}(t) \\ \vdots \\
		y_{\tau^i_{M_i}}(t) \end{pmatrix} \rightarrow 0
	\]
	and since $L_i$ is strongly connected we find:
	\[
	y_{\tau^i_a}-y_{\tau^i_b}(t) \rightarrow 0
	\]
	as $t\rightarrow \infty$ which implies the first bullet point of the
	theorem if we set $y^i_s=y^i_{\tau^i_1}$.
	
	Next we consider an agent not part of any of the basic bicomponents.
	Define the following matrix:
	\begin{equation}\label{Bdef}
		B= \begin{pmatrix}
			\beta_{11} M_1^{-1} \textbf{1}_{M_1}\T & \quad\cdots\quad &
			\beta_{1k} M_k^{-1} \textbf{1}_{M_k}\T \\
			\vdots & & \vdots \\
			\beta_{M_01} M_1^{-1} \textbf{1}_{M_1}\T & \quad\cdots\quad &
			\beta_{1k} M_k^{-1} \textbf{1}_{M_k}\T 
		\end{pmatrix}
	\end{equation}
	Then it is easy to verify from the structure of the kernel of $L$
	presented in \eqref{betamatrix} that
	\[
	\Ker L \subset \ker \begin{pmatrix} -I & B \end{pmatrix}.
	\]
	which implies that there exists a matrix $V$ such that
	\begin{equation}\label{VLdef}
		VL=\begin{pmatrix} -I & B \end{pmatrix}.
	\end{equation}
	but then weak synchronization implies that
	\[
	(I \otimes VL) y(t)\rightarrow 0
	\]
	Using \eqref{Bdef} and \eqref{VLdef} this implies:
	\begin{multline*}
		y_j(t) - \beta_{j1} M_1^{-1} \sum_{j=1}^{M_1} y_{\tau^1_j}(t)
		- \beta_{j2} M_2^{-1} \sum_{j=1}^{M_2} y_{\tau^2_j}(t) - \\
		\cdots
		- \beta_{jk} M_k^{-1} \sum_{j=1}^{M_k} y_{\tau^k_j}(t) \rightarrow 0
	\end{multline*}
	as $t\rightarrow \infty$ for $j=1,\ldots,M_0$. Using that
	$y_{\tau^i_j}(t)-y^i_s(t)\rightarrow 0$ established earlier this
	yields:
	\[
	y_j(t) - \beta_{j1} y_{\tau^1_s}(t)
	- \beta_{j2} y_{\tau^2_s}(t) - \cdots
	- \beta_{jk} y_{\tau^k_s}(t) \rightarrow 0
	\]
	as $t\rightarrow \infty$ for $j=1,\ldots,M_0$ which yields the
	second bullet point of the theorem.
\end{proof}

\section{The scale-free protocol for output synchronization and weak synchronization of heterogeneous MAS}

Since it is known that many multi-agent systems suffer from scale
fragility it is desirable if our protocol \eqref{protocol} for the
agent \eqref{system} to be scale free. This implies that the protocol
\eqref{protocol} for agent $i$ must be designed only based on agent
model $i$.  This is formally defined below:

\begin{definition}[\textbf{Scale-free output synchronization}]\label{def}
	The family of protocols \eqref{protocol} is said to achieve
	scale-free output synchronization for the family of agents
	\eqref{system} if the following property holds.
	
	For any selection of agents $\{\tau_1,\ldots, \tau_M\}$ and for any
	associated graph $\mathcal{G}\in \mathbb{G}^M$ which has a directed spanning tree
	and $M$ nodes (with associated Laplacian $\tilde{L}$), we have that
	\[
	\begin{system*}{ccl}
		{x}_{s,e}^+ &=& [\tilde{A}_s + \tilde{B}_s (\tilde{L} \otimes \tilde{H}_s)] x_{s,e} \\
		y_s &=& \tilde{C}_s x_{s,e}
	\end{system*}
	\]
	achieves synchronization, i.e.
	\[
	y_i(t)-y_j(t) \rightarrow 0
	\]
	as $t\rightarrow \infty$ for any $i,j\in \{\tau_1,\ldots, \tau_M \}$ and all
	possible initial conditions for $\tilde{x}_{s,e}$ where we define
	\begin{align*}
		\tilde{A}_s &=\diag\{\tilde{A}_{\tau_1},\ldots,\tilde{A}_{\tau_M}\},\qquad
		\tilde{B}_s =\diag\{\tilde{B}_{\tau_1},\ldots,\tilde{B}_{\tau_M}\}, \\
		\tilde{C}_s &=\diag\{\tilde{C}_{\tau_1},\ldots,\tilde{C}_{\tau_M}\},\qquad
		\tilde{H}_s =\diag\{\tilde{H}_{\tau_1},\ldots,\tilde{H}_{\tau_M}\}
	\end{align*}
	and
	\[
	x_{s,e} = \begin{pmatrix} x_{e,\tau_1} \\ \vdots \\ x_{e,\tau_M} \end{pmatrix}\qquad
	y_s = \begin{pmatrix} y_{\tau_1} \\ \vdots \\ y_{\tau_M} \end{pmatrix},
	\]
\end{definition}

\begin{remark}
	For heterogeneous MAS, when all agents are introspective, scale-free
	collaborative protocols have been designed in
	\cite{donya-liu-saberi-stoorvogel-acc2020}. When all agents are
	non-introspective and passive, static scale-free non-collaborative
	protocols have been designed for strongly connected graph in
	\cite{stoorvogel-saberi-liu-nojavanzadeh-ijrnc19}.
\end{remark}

Scale free designs are used in the context that the network is not
known. But this makes it hard to verify the assumption that the
network has a directed spanning tree since verifying this
intrinsically requires knowledge of the network. In that sense, the
concept of scale-free weak synchronization is more appropriate:

\begin{definition}[\textbf{Scale-free
		weak synchronization}]\label{def2}
	The family of protocols \eqref{protocol} is said to achieve scale-free
	weak synchronization for the family of agents \eqref{system} if the
	following property holds.
	
	For any selection of agents $\{\tau_1,\ldots, \tau_M\}$ and for any
	associated graph $\mathcal{G}\in \mathbb{G}^M$ (with associated
	Laplacian $\tilde{L}$), we have that
	\[
	\begin{system*}{ccl}
		{x}_{s,e}^+ &=& [\tilde{A}_s + \tilde{B}_s (\tilde{L} \otimes \tilde{H}_s)] x_{s,e} \\
		y_s &=& \tilde{C}_s x_{s,e}
	\end{system*}
	\]
	(using the same notation as in Definition \ref{def} achieves weak
	synchronization, i.e.
	\[
	\zeta_i(t) \rightarrow 0
	\]
	as $t\rightarrow \infty$ for any $i\in \{\tau_1,\ldots, \tau_M \}$ and all
	possible initial conditions for $\tilde{x}_{s,e}$ 
\end{definition}

The next objective of this paper is to show that protocols that achieve scale-free output synchronization as defined in Definition \ref{def},  also achieve weak
synchronization in the absence of a spanning tree due to a fault.

\begin{theorem}\label{thm_f1heter-c}
	Consider a continuous-time heterogeneous MAS with agent dynamics
	\eqref{system} with protocols of the form \eqref{protocol}.
	
	In that case, we achieve scale-free output synchronization (as
	defined in Definition \ref{def}) if and only if we achieve
	scale-free weak synchronization (as
	defined in Definition \ref{def2}).
\end{theorem}

\begin{remark}
	Note that scale-free weak synchronization implies scale-free output
	synchronization even if we allow for nonlinear protocols. However,
	our proof below explicitly depends on linearity for the reverse
	implication. We can for instance easily obtain extensions of the
	above theorems if we have protocols containing time-delays because
	that preserves the linearity.
\end{remark}

\begin{proof}
	From Lemma \ref{xxxxx2} we obtain that weak synchronization implies output
	synchronization if the network contains a directed spanning
	tree. This immediately implies that scale-free weak synchronization
	implies scale-free output synchronization.
	
	Remains to establish that we obtain scale-free weak synchronization
	if we know that we have achieved scale-free output
	synchronization. If we look at the interconnection of \eqref{system}
	and \eqref{protocol} we can write this of the form:
	\begin{equation}\label{clsystem}
		\begin{system*}{ccl}
			x_{e,i}^+ &=& \tilde{A}_ix_{e,i} + \tilde{B}_i \tilde{\zeta}_i \\
			y_i &=&  \tilde{C}_i x_{e,i}\\
			z_i &=&  \tilde{H}_i x_{e,i}
		\end{system*}
	\end{equation}
	with
	\begin{equation}
		\tilde{\zeta}_i = \sum_{j=1}^N \ell_{i,j} z_j
	\end{equation} 
	\begin{align*}
		\tilde{A}_i &=\begin{pmatrix}
			A_i&B_iM_i\\
			L_{i,m}C_{i,m}&K_i
		\end{pmatrix}, \tilde{B}_i=\begin{pmatrix}
			0&0\\
			L_{i}&L_{i,e}
		\end{pmatrix}, \\
		\tilde{C}_i &=\begin{pmatrix}
			C_i&0
		\end{pmatrix}, \tilde{H}_i=\begin{pmatrix}
			C_i&0\\
			0&N_i
		\end{pmatrix}
	\end{align*}
	If we define
	\begin{align*}
		\tilde{A} &=\diag\{\tilde{A}_1,\ldots,\tilde{A}_N\},\qquad
		\tilde{B} =\diag\{\tilde{B}_1,\ldots,\tilde{B}_N\}, \\
		\tilde{C} &=\diag\{\tilde{C}_1,\ldots,\tilde{C}_N\},\qquad
		\tilde{H}=\diag\{\tilde{H}_1,\ldots,\tilde{H}_N\}
	\end{align*}
	and
	\[
	x_e = \begin{pmatrix} x_{e,1} \\ \vdots \\ x_{e,N} \end{pmatrix}\qquad
	y = \begin{pmatrix} y_1 \\ \vdots \\ y_N \end{pmatrix},
	\]
	then we can write the complete system as:
	\begin{equation}\label{clsysteme}
		\begin{system*}{ccl}
			{x}_e^+ &=& [\tilde{A} + \tilde{B}  (L \otimes \tilde{H})] x_e \\
			y &=& \tilde{C} x_e
		\end{system*}
	\end{equation}
	
	We next consider the following $k$ differential equations:
	\begin{equation}\label{clsystemei}
		\begin{system*}{ccl}
			(x^i_e)^+ &=& [\tilde{A} + \tilde{B}  (L \otimes \tilde{H})] x^i_e \\
			y^i &=& \tilde{C} x_e^i
		\end{system*}
	\end{equation}
	for $i=1,\ldots , k$ where
	\[
	x^i_e = \begin{pmatrix} x^i_{e,1} \\ \vdots \\ x^i_{e,N} \end{pmatrix}\qquad
	y^i = \begin{pmatrix} y_1^i \\ \vdots \\ y^i_N \end{pmatrix}.
	\]
	Each of these systems is kind of connected to one of the basic
	bicomponents through its initial conditions. In particular, assume
	agent $j$ is part of a basic bicomponent $i$ then we choose
	\begin{equation*}
		\begin{system}{ll}
			x^v_{e,j}(0)=x_{e,j}(0), & \qquad v=i,\\
			x^v_{e,j}(0)=0, & \qquad v\neq i,
		\end{system}
	\end{equation*}
	and hence
	\[
	x^1_{e,j}(0)+\cdots+x^k_{e,j}(0)=x_{e,j}(0).
	\]
	
	On the other hand, if agent $j$ is not part of any basic
	bicomponent, then there is at least one bicomponent $i$ from which
	agent $j$ can be reached.
	
	Assume that $i$ is the bicomponent with smallest index with this
	property (which implies $\beta_{j,i}\neq 0$). then we choose
	\begin{equation*}
		\begin{system}{ll}
			x^v_{e,j}(0)=x_{e,j}(0), & \qquad v=i,\\
			x^v_{e,j}(0)=0, & \qquad v\neq i,
		\end{system}
	\end{equation*}
	and hence again
	\[
	x^1_{e,j}(0)+\cdots+x^k_{e,j}(0)=x_{e,j}(0).
	\]
	
	In other words, we have
	\[
	\begin{pmatrix}
		x^1_{e,1}(0)+\cdots+x^k_{e,1}(0)\\
		x^1_{e,2}(0)+\cdots+x^k_{e,2}(0)\\
		\vdots\\
		x^1_{e,N}(0)+\cdots+x^k_{e,N}(0)
	\end{pmatrix}
	=\begin{pmatrix}
		x_{e,1}(0)\\
		x_{e,2}(0)\\
		\vdots\\
		x_{e,N}(0)
	\end{pmatrix}
	\]
	It means that we have
	\[
	x^1_{e}(0)+\cdots+x^k_{e}(0)=x_e(0)
	\]
	with 
	\[
	x_e(0)=\begin{pmatrix}
		x_{e,1}(0)\\
		x_{e,2}(0)\\
		\vdots\\
		x_{e,N}(0)
	\end{pmatrix}
	\]
	Using equation \eqref{clsystemei} we obtain
	\begin{equation} \label{starty}
		\begin{system*}{l}
			(x^1_{e}+\cdots+{x}^k_{e})^+= [\tilde{A} + \tilde{B}  (L
			\otimes \tilde{H})]  (x^1_{e}+\cdots+x^k_{e})\\ 
			(y^1+\cdots+y^n)=\tilde{C}(x^1_{e}+\cdots+x^k_{e})
		\end{system*}
	\end{equation}
	But from equation \eqref{clsysteme} and \eqref{starty} we see
	that $x^1_{e}+\cdots+x^k_{e}$ and $x_e$ satisfy the same
	differential equation and have the same initial condition. It
	implies that
	\begin{equation}\label{sum}
		x_e=x^1_e+\cdots+x^k_e,\qquad y=y^1+\cdots+y^k
	\end{equation}
	Next consider $x^i_e$. Define:
	\[
	\begin{pmatrix} \gamma^i_1 \\ \vdots \\ \gamma^i_N \end{pmatrix}
	\]
	as the $i$'th column of the matrix \eqref{betamatrix}. This implies
	\begin{equation} \label{gammasum}
		\sum_{v=1}^N \ell_{j,v} \gamma^i_v = 0
	\end{equation}
	Consider an agent $j$ for which $\gamma^i_j=0$. Then we note that all
	terms of the summation are nonpositive since the $\gamma^i_v$ are
	nonnegative and the $\ell_{j,v}$ for $v\neq j$ are nonpositive. This
	implies:
	\begin{equation}\label{14a}
		\ell_{j,v} \gamma^i_v = 0 \quad \text{ for } v=1,\ldots, N.
	\end{equation}
	This yields that an agent $j$ for which $\gamma_j=0$ only depends on
	other agents $v$ (i.e. $\ell_{j,v}\neq 0$) for which $\gamma_v=0$. But since all
	agents for which $\gamma_v=0$ satisfy, by construction, $x^i_{e,j}(0)=0$
	this yield that $x^i_{e,j}(t)=0$ for all $t>0$.
	
	We consider all agents  $\tau^i_1,\ldots,\tau^i_{N_i}$ for which the
	corresponding $\gamma_v$ is nonzero, i.e.
	\begin{equation}\label{gggH}
		\gamma_{\tau^i_j} \neq 0 \quad \text{ for } j=1,\ldots , N_i
	\end{equation}
	and let $L_i$ denote the matrix obtained from $L$ by deleting both
	columns and rows whose index is not contained in the set
	$\tau^i_1,\ldots,\tau^i_{N_i}$. It can be shown that this
	matrix has rank $N_i-1$. After all, we already know
	\begin{equation} \label{subvector}
		\begin{pmatrix}
			\gamma_{\tau^i_1} \\ \vdots \\  
			\gamma_{\tau^i_{N_i}} 
		\end{pmatrix}
	\end{equation}
	is contained in the null space of $L_i$. If, additionally,
	\[
	\begin{pmatrix}
		\eta_{\tau^i_1} \\ \vdots \\  
		\eta_{\tau^i_{N_i}} 
	\end{pmatrix}
	\]
	is in the null space of $L_i$ and linearly independent of
	\eqref{subvector} then it is easily verified that
	\[
	\eta=\begin{pmatrix}
		\eta_1 \\ \vdots \\ \eta_N 
	\end{pmatrix}
	\]
	is in the null space of $L$ where we have chosen $\eta_i=0$ when $i
	\not\in \{ \tau^i_1, \ldots, \eta_{\tau^i_{N_i}} \}$. Here we use
	that \eqref{14a} implies that
	\[
	\ell_{j,v} \eta_v = 0 \quad \text{ for } v=1,\ldots N.
	\]
	for any $j$ for which $\gamma_j=0$ (recall that $\ell_{j,v}\neq 0$
	implies $\gamma_v\neq 0$). But given the structure of the null space
	of $L$ given by \eqref{betamatrix}, $L\eta=0$ yields a
	contradiction.
	
	Note that $L_i$ has the structure of a Laplacian except for the zero
	row sum.  However, the matrix:
	\[
	\Gamma = \diag\{  \gamma_{\tau^i_1}, \ldots ,
	\gamma_{\tau^i_{N_i}}\}
	\]
	is invertible and
	\[
	\tilde{L}_i = \Gamma^{-1} L_i \Gamma
	\]
	is a classical Laplacian matrix with zero row sum. Moreover, it
	contains a directed spanning tree since its rank is equal to $N_i-1$.
	
	We consider agents $\tau^i_1,\ldots,\tau^i_{N_i}$. Using the above,
	we obtain that
	\begin{equation}\label{clsystemei2}
		\begin{system*}{ccl}
			x_{s,e}^+ &=& [\tilde{A}_s + \tilde{B}_s  (L_i \otimes \tilde{H})] x_{s,e} \\
			y_s &=& \tilde{C} x_{s,e}
		\end{system*}
	\end{equation}
	where
	\begin{align*}
		\tilde{A}_s &=\diag\{\tilde{A}_{\tau^i_1},\ldots,\tilde{A}_{\tau^i_{N_i}}\},\qquad
		\tilde{B}_s =\diag\{\tilde{B}_{\tau^i_1},\ldots,\tilde{B}_{\tau^i_{N_i}}\}, \\
		\tilde{C}_s &=\diag\{\tilde{C}_{\tau^i_1},\ldots,\tilde{C}_{\tau^i_{N_i}}\},\qquad
		\tilde{H}_s =\diag\{\tilde{H}_{\tau^i_1},\ldots,\tilde{H}_{\tau^i_{N_i}}\}
	\end{align*}
	and
	\[
	x_{s,e} = \begin{pmatrix} x^i_{e,\tau^i_1} \\ \vdots \\
		x^i_{e,\tau^i_{N_i}} \end{pmatrix}\qquad 
	y_s = \begin{pmatrix} y^i_{\tau^i_1} \\ \vdots \\ y^i_{\tau^i_{N_i}} \end{pmatrix}.
	\]
	But then we obtain:
	\begin{equation}\label{clsystemei3}
		\begin{system*}{ccl}
			\tilde{x}_{s,e}^+ &=& [\tilde{A}_s + \tilde{B}_s  (\tilde{L}_i
			\otimes \tilde{H}_s)] \tilde{x}_{s,e} \\ 
			\tilde{y}_s &=& \tilde{C}_s \tilde{x}_{s,e}
		\end{system*}
	\end{equation}
	where $\tilde{x}_{s,e}=(\Gamma^{-1}\otimes I) x_{s,e}$ and
	$\tilde{y}_s=(\Gamma^{-1}\otimes I) y_s$. Since our protocol achieved
	scale-free synchronization and the network associated to
	$\tilde{L}_i$ contains a directed spanning tree we obtain
	output synchronization using Definition \ref{def} and therefore also
	weak synchronization by Lemma \ref{xxxxx}, i.e.
	\[
	(\tilde{L}_i \otimes I) \tilde{y}_s \rightarrow 0
	\]
	as $t\rightarrow \infty$ which implies
	\[
	(L_i\otimes I) y_s \rightarrow 0
	\]
	Given the way we constructed $y_s$, this implies:
	\[
	(L\otimes I) y^i \rightarrow 0
	\]
	Since this is true for $i=1,\ldots, k$ we can use \eqref{sum} to
	establish:
	\[
	(L\otimes I) y \rightarrow 0
	\]
	In other words, we achieve weak synchronization since the above
	derivation is valid for all possible initial conditions.
\end{proof}

\section{Numerical examples}

In this section, we consider a special case of protocol
\eqref{protocol}: all agent models are introspective and protocols are
collaborative. We choose the existing examples in both continuous- and discrete-time presented in
\cite{donya-liu-saberi-stoorvogel-acc2020} and \cite{wang-saberi-yang}.

\subsection{Continuous-time case}
We consider agent models of the form \eqref{system} with the following three groups of parameters.
For Model 1 we have:
\begin{equation*}
	A_i=\begin{pmatrix}
		0&1&0&0\\0&0&1&0\\0&0&0&1\\0&0&0&0
	\end{pmatrix}, B_i=\begin{pmatrix}
		0&1\\0&0\\1&0\\0&1
	\end{pmatrix}, C_i\T=\begin{pmatrix}
		1\\0\\0\\0
	\end{pmatrix},
	(C_i^m)\T=\begin{pmatrix}
		1\\1\\0\\0
	\end{pmatrix}
\end{equation*}
while for Model 2 we have:
\begin{equation*}
	A_i=\begin{pmatrix}
		0&1&0\\0&0&1\\0&0&0
	\end{pmatrix},B_i=\begin{pmatrix}
		0\\0\\1
	\end{pmatrix},C_i\T=\begin{pmatrix}
		1\\0\\0
	\end{pmatrix},
	(C_i^m)\T=\begin{pmatrix}
		1\\1\\0
	\end{pmatrix}.
\end{equation*}
Finally, for Model 3 we have:
\begin{align*}
	&A_i=\begin{pmatrix}
		-1&0&0&-1&0\\0&0&1&1&0\\0&1&-1&1&0\\0&0&0&1&1\\-1&1&0&1&1
	\end{pmatrix},\quad B_i=\begin{pmatrix}
		0&0\\0&0\\0&1\\0&0\\1&0
	\end{pmatrix},\\
	&C_i=\begin{pmatrix}
		0&0&0&1&0
	\end{pmatrix},\quad {C_i^m}=\begin{pmatrix}
		1&1&0&0&0
	\end{pmatrix}
\end{align*}
For the protocols we use the design methodology of
\cite{donya-liu-saberi-stoorvogel-acc2020}, we first choose a target
model:
\[
A=\begin{pmatrix}
	0&1&0\\0&0&1\\0&-1&0
\end{pmatrix}, B=\begin{pmatrix}0\\0\\1\end{pmatrix}, C=\begin{pmatrix}1&0&0\end{pmatrix}.
\] 
Then we assign precompensators (for each different model) such that
the behavior of the system combining model with precompensator
approximately behaves as the target model. Finally we combine
this precompensator (which is different for each model) with a
homogeneous protocol designed for the target model. For details we
refer to \cite{donya-liu-saberi-stoorvogel-acc2020}.

\begin{figure}[ht]
	\includegraphics[width=8cm]{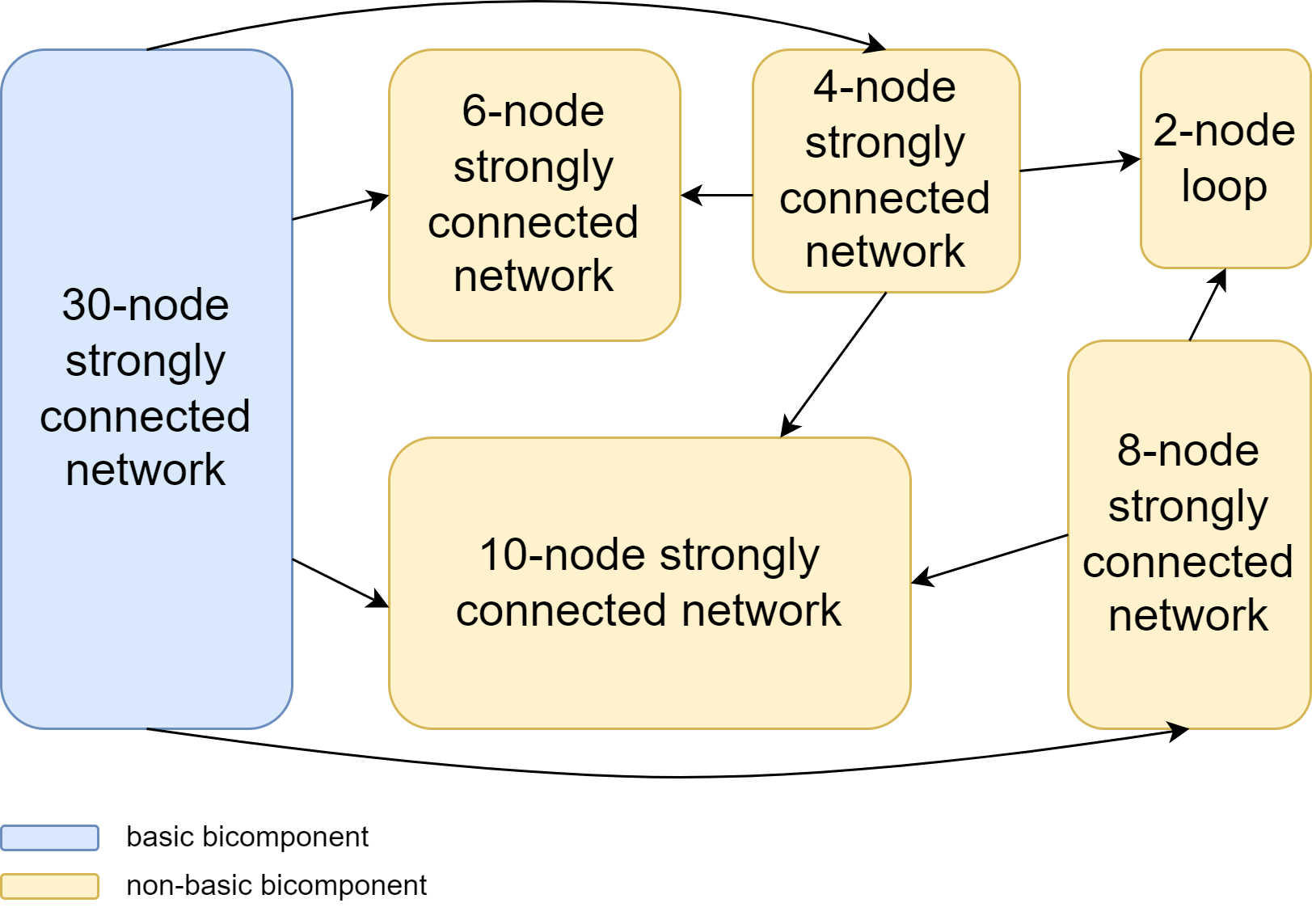}
	\centering
	\caption{The 60-nodes communication network with spanning tree.}\label{f5}
\end{figure}

We consider scale-free output synchronization result for the 60-node
heterogeneous network shown in figure \ref{f5}, which contains a
directed spanning tree. For this example, each agent is randomly
assigned one of the above four models.

\begin{figure}[ht]
	\includegraphics[width=8cm]{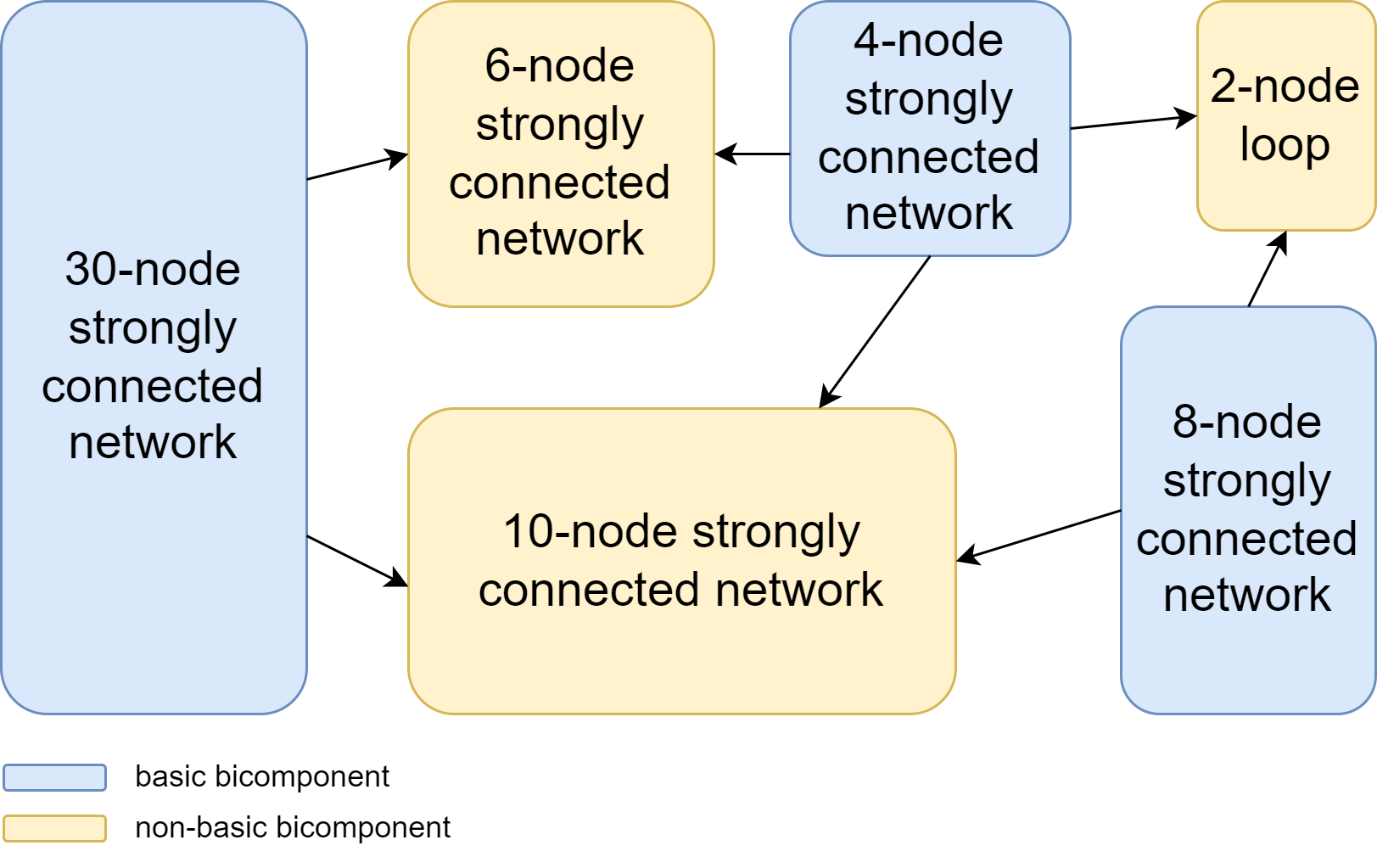}
	\centering
	\caption{The communication network without spanning tree. The
		links are broken due to faults. }\label{f4}  
\end{figure}

When some links have faults, the communication network
might lose its directed spanning tree. For example, if two specific
links are broken in the original 60-node network given by Figure
\ref{f5}, then we obtain the network as given in Figure \ref{f4}

It is obvious that there is no spanning tree in Figure \ref{f4}. We
obtain three basic bicomponents (indicated in blue): one containing 30
nodes, one containing 8 nodes and one containing 4 nodes. Meanwhile,
there are three non-basic bicomponents: one containing 10 nodes, one
containing 6 nodes and one containing 10 nodes, which are indicated in
yellow.

\begin{figure}[ht]
	\includegraphics[width=9cm]{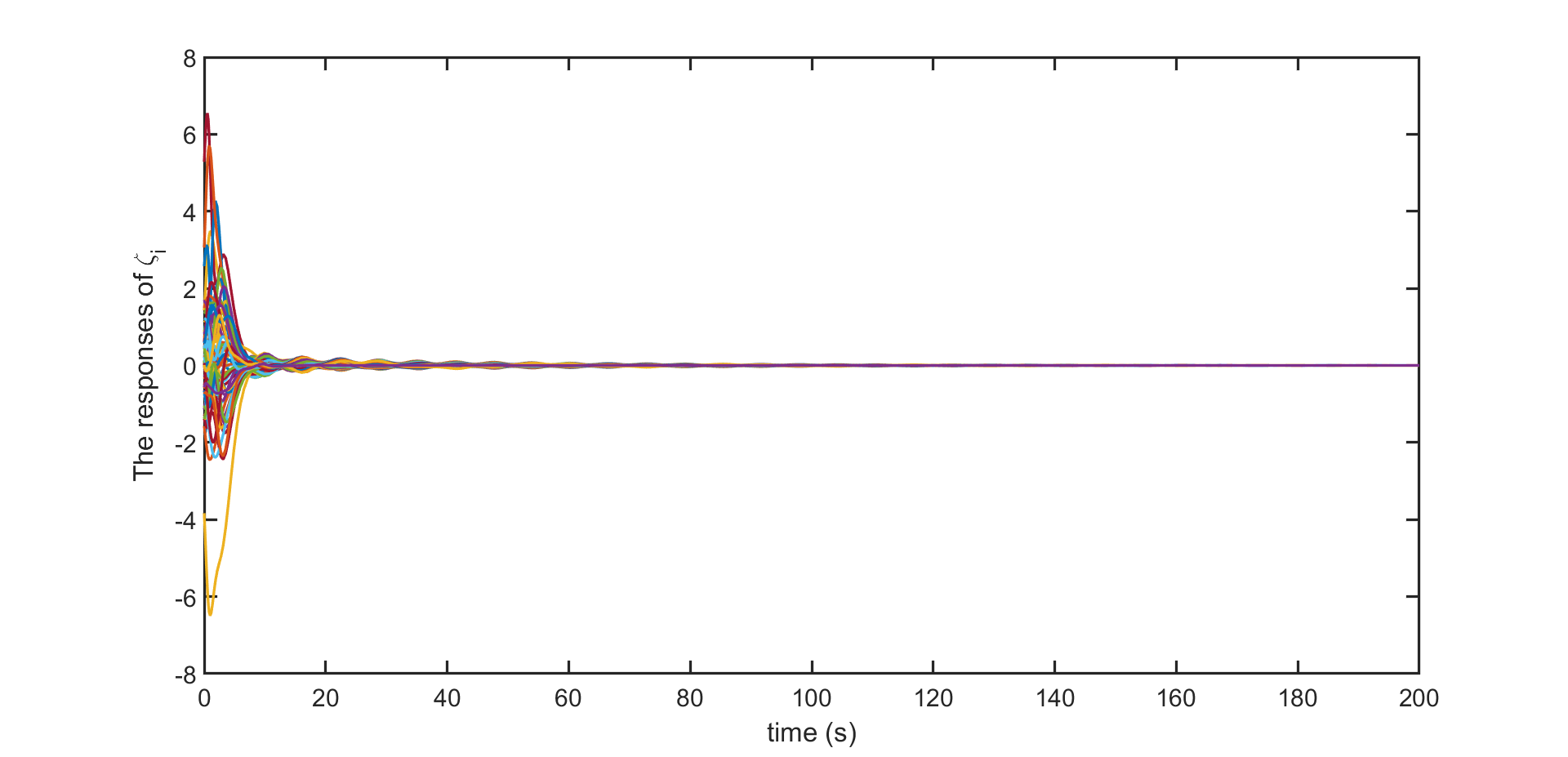}
	\centering
	\caption{The trajectory of $\zeta_i$ for continuous-time MAS.}\label{zeta60n}
\end{figure}

By using the scale-free protocol, we obtain $\zeta_i\to 0$ as
$t\to \infty$, which means weak synchronization is achieved in the
absence of connectivity, see Fig. \ref{zeta60n}. It implies that the
available network data for each agent goes to zero and the communication
network becomes inactive.

\begin{figure}[t]
	\includegraphics[width=9cm]{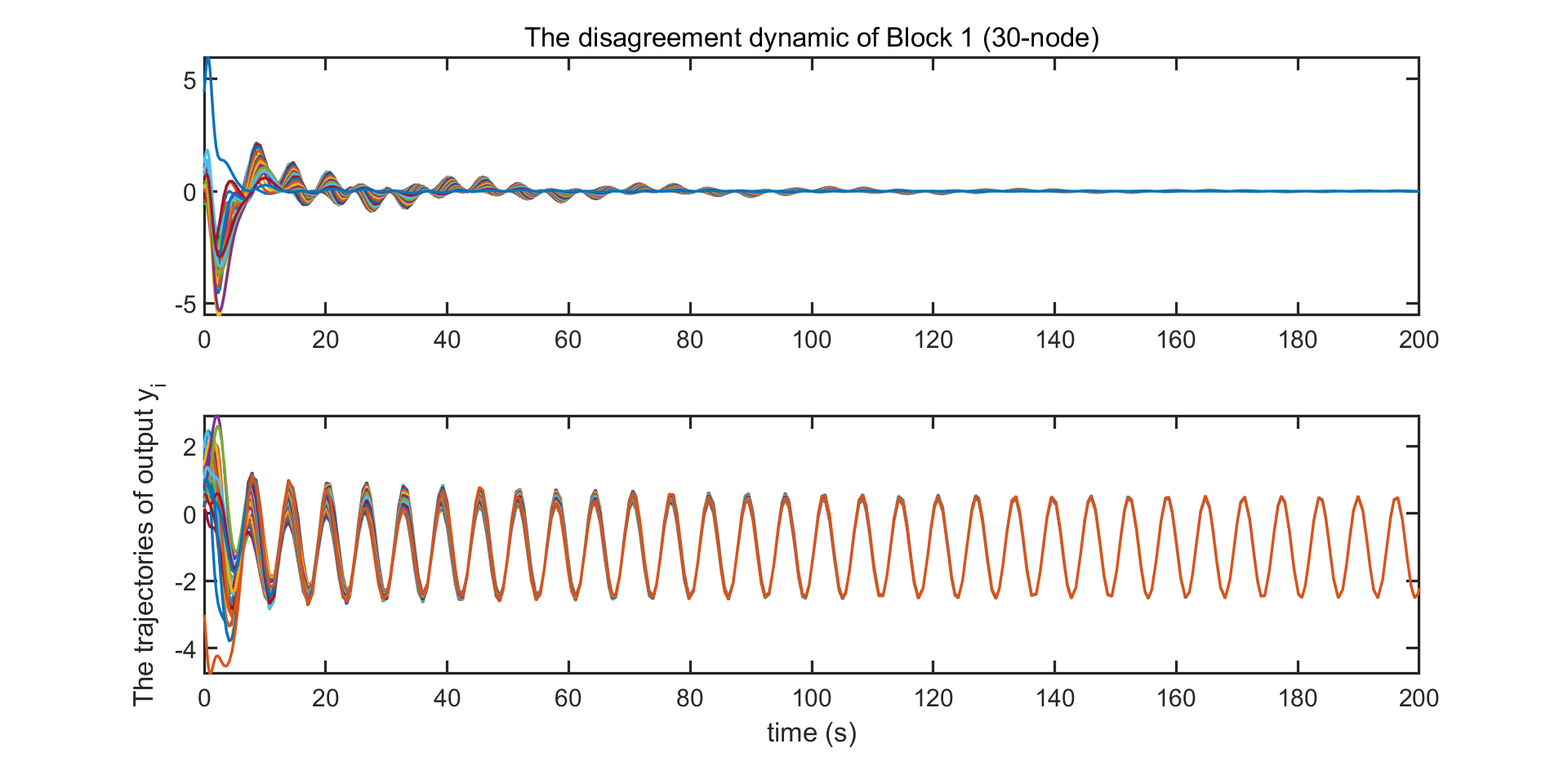}
	\centering
	\caption{Basic bicomponent 1 for continuous-time MAS: disagreement dynamic among the
		agents and synchronized output trajectories.}\label{ct30co}
\end{figure}

\begin{figure}[t]
	\includegraphics[width=9cm]{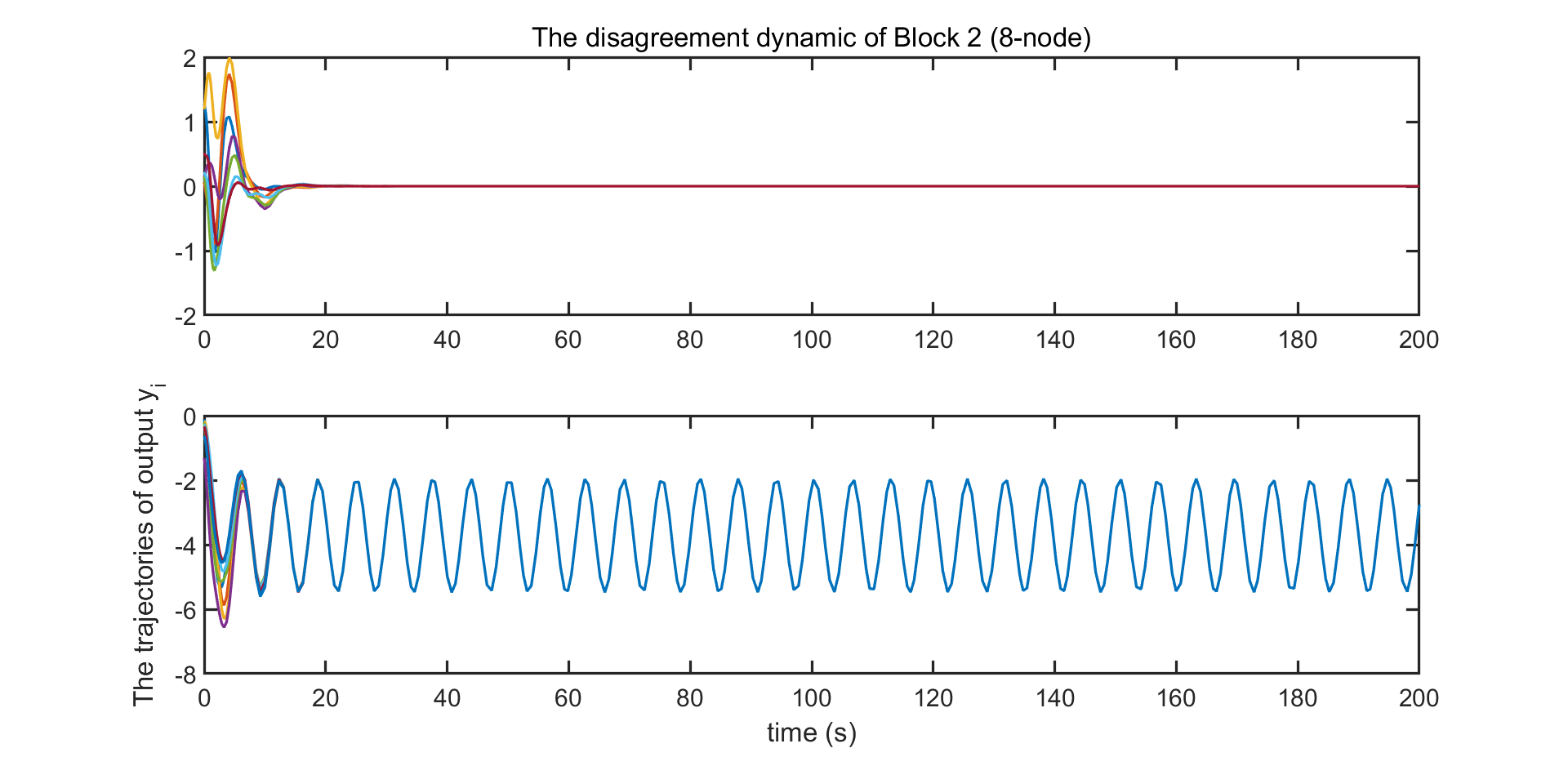} \centering
	\caption{Basic bicomponent 2 for continuous-time MAS: disagreement dynamic among the
		agents and synchronized output trajectories.}\label{ct8co}
\end{figure}

\begin{figure}[t]
	\includegraphics[width=9cm]{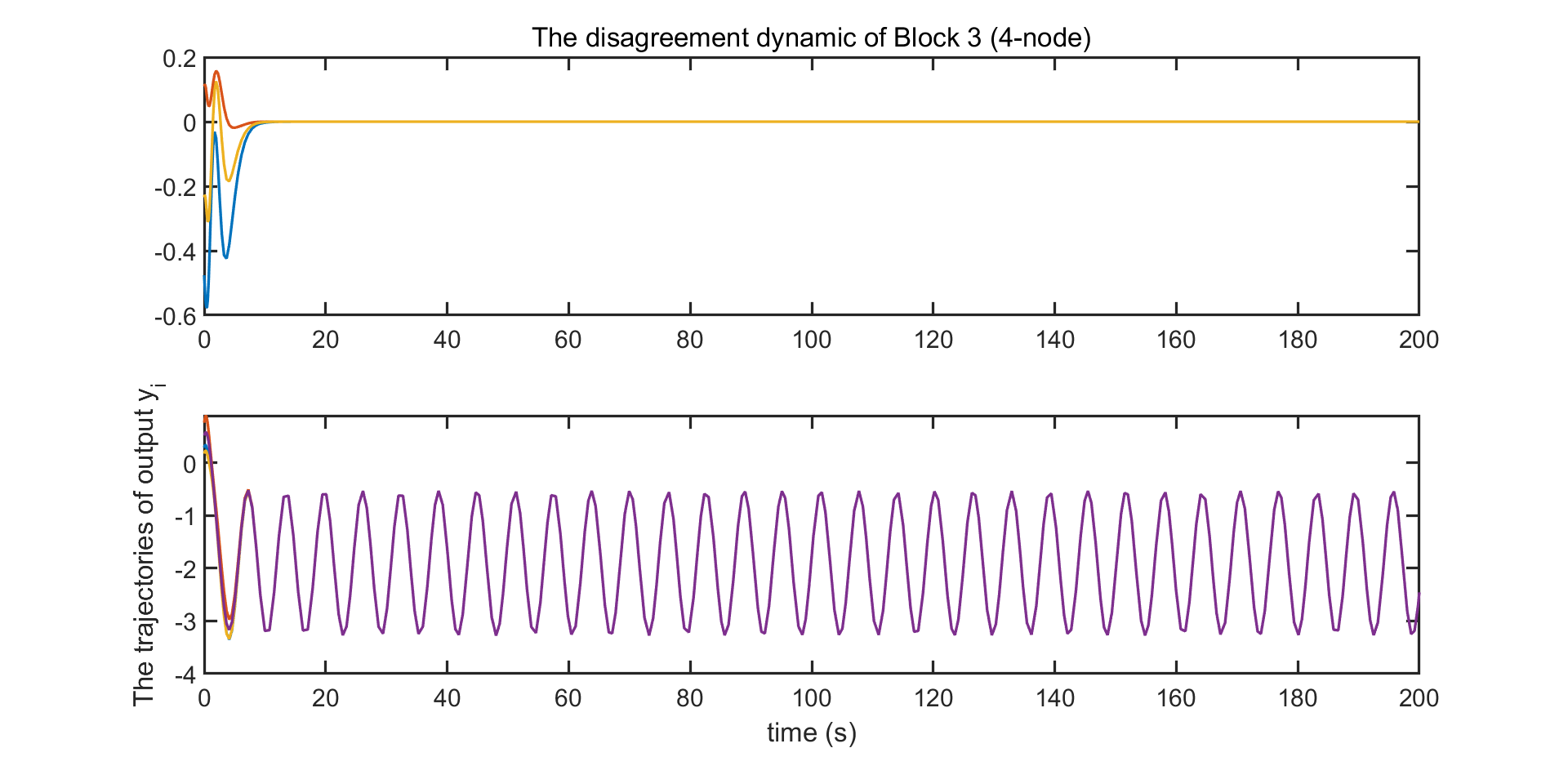}
	\centering
	\caption{Basic bicomponent 3 for continuous-time MAS: disagreement dynamic among the
		agents and synchronized output trajectories.}\label{ct1colead}
\end{figure}

We have seen that for the 60-node network given in Figure \ref{f5}
this protocol indeed achieves output synchronization. If we apply the
same protocol to the network described by Figure \ref{f4} which does
not contain a directed spanning tree, we again consider the six
bicomponents constituting the network. We see that, consistent with
the theory, we get output synchronization within the three basic
bicomponents as illustrated in Figures \ref{ct30co}, \ref{ct8co} and
\ref{ct1colead} respectively. Clearly, the disagreement dynamic among
the agents (the errors between the output of agents) goes to zero
within each basic bicomponent. According to Theorem \ref{thm_dy}, we
obtain that any agent outside of the basic bicomponent converge to a
convex combination of the synchronized trajectories of the basic
bicomponents, i.e., agents in either one of these non-basic
bicomponents converge to a convex combination of the synchronized
trajectories of the basic bicomponents 1, 2 and 3. Note that we do not
necessarily have that all agents within a specific non-basic
bicomponent converge to the same asymptotic behaviour.


\subsection{Discrete-time case}

Consider discrete-time agent models of the form \eqref{system} with
four different sets of parameters
\begin{equation*}
	A_i=\begin{pmatrix}
		0&1&0&0\\0&0&1&0\\-1&0&0&-1\\0&-1&0&0
	\end{pmatrix}, B_i=\begin{pmatrix}
		0&0\\0&0\\0&1\\1&0
	\end{pmatrix}, C_i\T=\begin{pmatrix}
		0\\0\\0\\1
	\end{pmatrix},
	{C_i^m}\T=\begin{pmatrix}
		0\\-1\\0\\1
	\end{pmatrix}
\end{equation*}
for model 1, and
\begin{equation*}
	A_i=\begin{pmatrix}
		0&1&0\\0&0&1\\0&0&0
	\end{pmatrix},B_i=\begin{pmatrix}
		0\\0\\1
	\end{pmatrix},C_i\T=\begin{pmatrix}
		1\\0\\0
	\end{pmatrix},
	{C_i^m}\T=\begin{pmatrix}
		1\\1\\0
	\end{pmatrix}
\end{equation*}
for model 2,
\begin{equation*}
	A_i=\begin{pmatrix}
		0&1\\0&0
	\end{pmatrix},B_i=\begin{pmatrix}
		0\\1
	\end{pmatrix},C_i\T=\begin{pmatrix}
		1\\0
	\end{pmatrix}, {C_i^m}\T=\begin{pmatrix}
		1\\1
	\end{pmatrix}
\end{equation*}
for model 3 and, finally,
\begin{equation*}
	A_i=\begin{pmatrix}
		0&1\\-2&-2
	\end{pmatrix},B_i=\begin{pmatrix}
		0\\1
	\end{pmatrix},C_i\T=\begin{pmatrix}
		1\\0
	\end{pmatrix}, {C_i^m}\T=\begin{pmatrix}
		1\\1
	\end{pmatrix}
\end{equation*}
for model 4. 

For the protocols we use the design methodology of
\cite{wang-saberi-yang} which is similar to the technique we used in
the continuous time. We first choose a target model:
\[
A=\begin{pmatrix}
	0&1&0\\0&0&1\\1&-2&2
\end{pmatrix}, B=\begin{pmatrix}0\\0\\1\end{pmatrix}, C=\begin{pmatrix}1&0&0\end{pmatrix}.
\]
We assign precompensators (for each different model) such that the
behavior of the system combining model with precompensator
approximately behaves as the target model. Finally, we combine this
precompensator (which is different for each model) with a homogeneous
protocol designed for the target model. For details, we refer to
\cite{wang-saberi-yang}.

As a result, we consider scale-free weak synchronization result for
the 60-node discrete-time heterogeneous network shown in Figure
\ref{f4}. Again each agent is randomly assigned one of these four
models in this case.

\begin{figure}[b]
	\includegraphics[width=9cm]{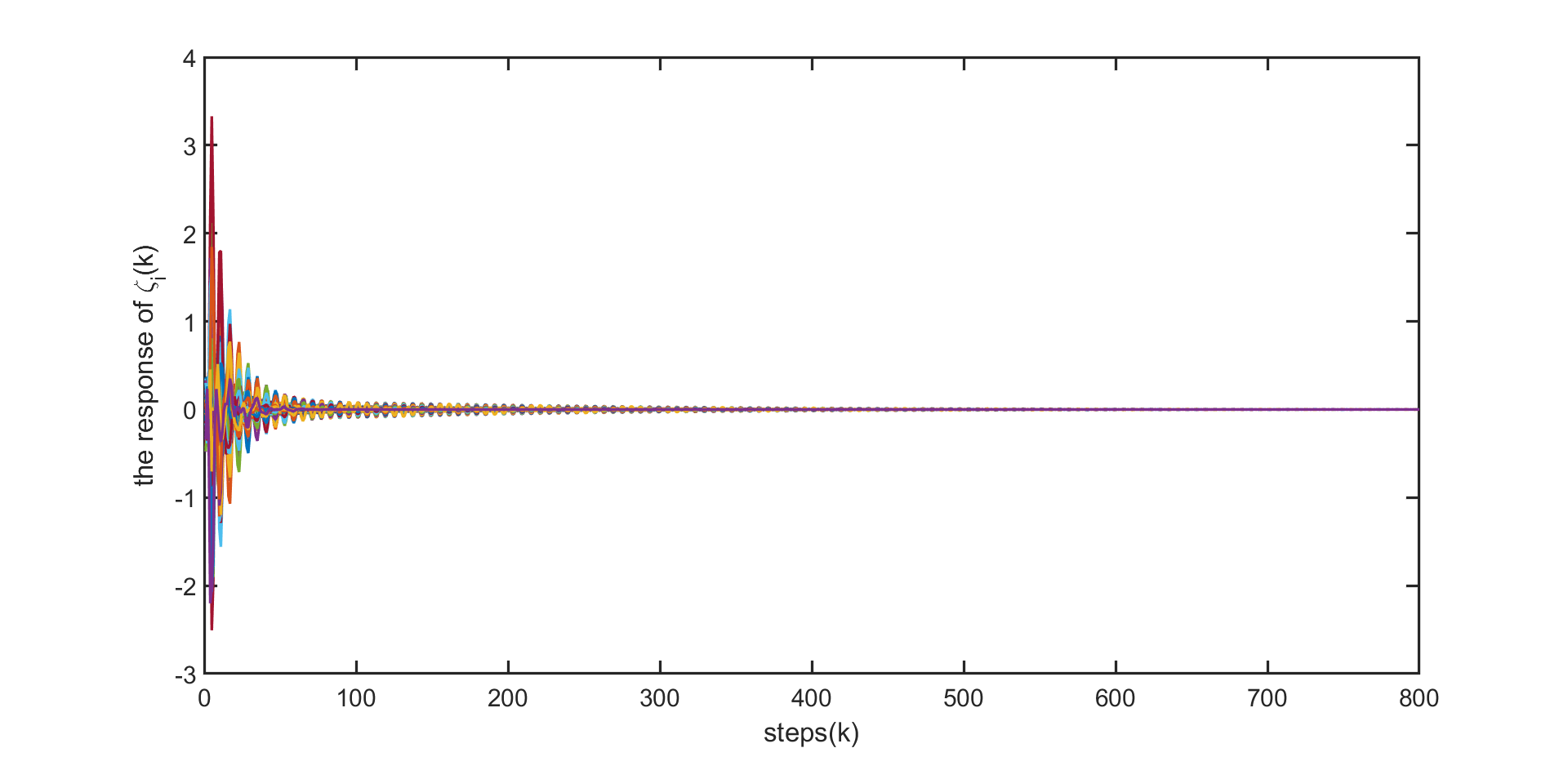}
	\centering
	\caption{The trajectory of $\zeta_i$ for discrete-time MAS.}\label{diszeta60n}
\end{figure}
\begin{figure}[ht]
	\includegraphics[width=9cm]{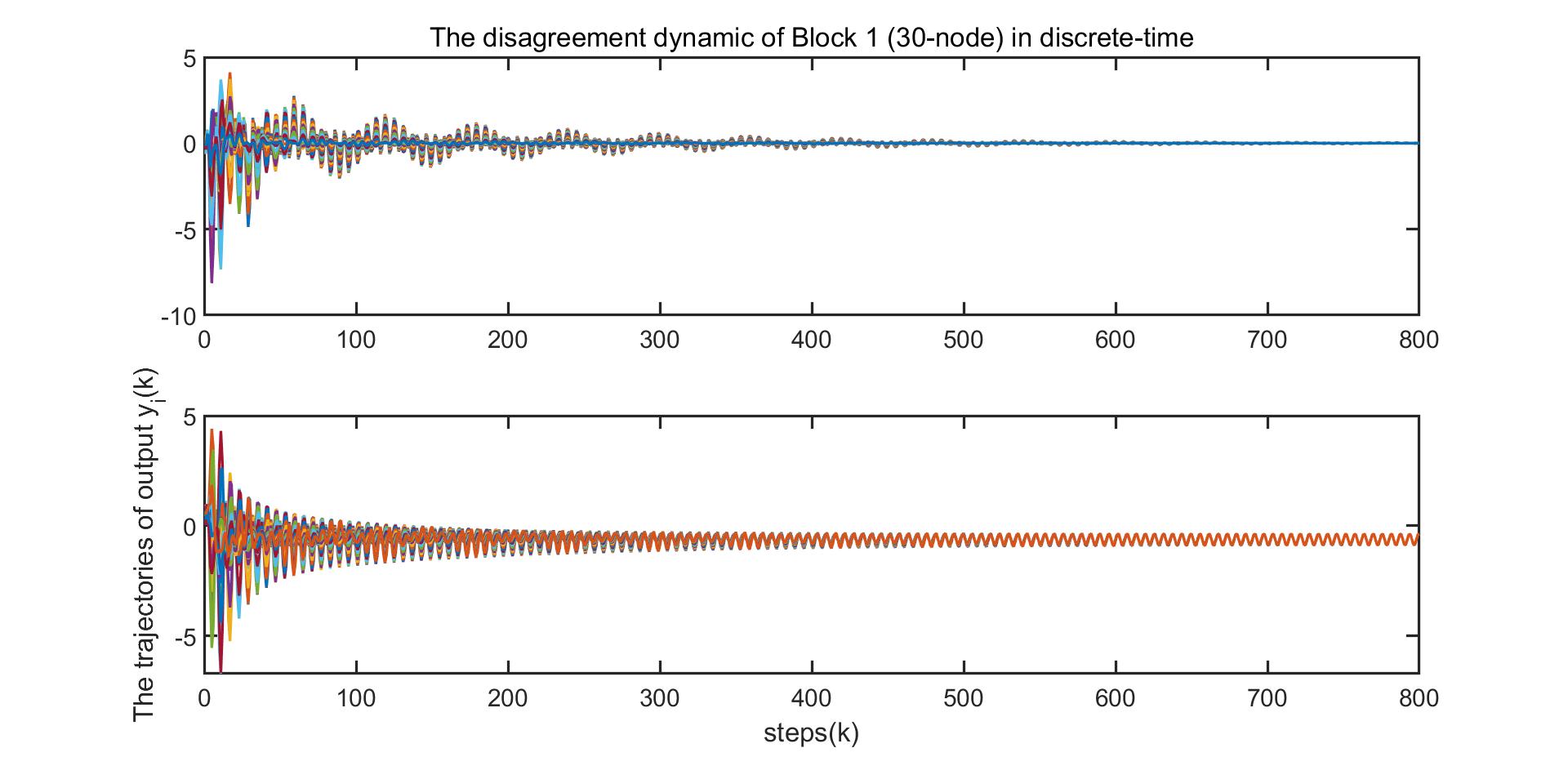}
	\centering
	\caption{Basic bicomponent 1 for discrete-time MAS: disagreement
		dynamic among the agents and synchronized output
		trajectories.}\label{dt30co} 
\end{figure}
\begin{figure}[ht]
	\includegraphics[width=9cm]{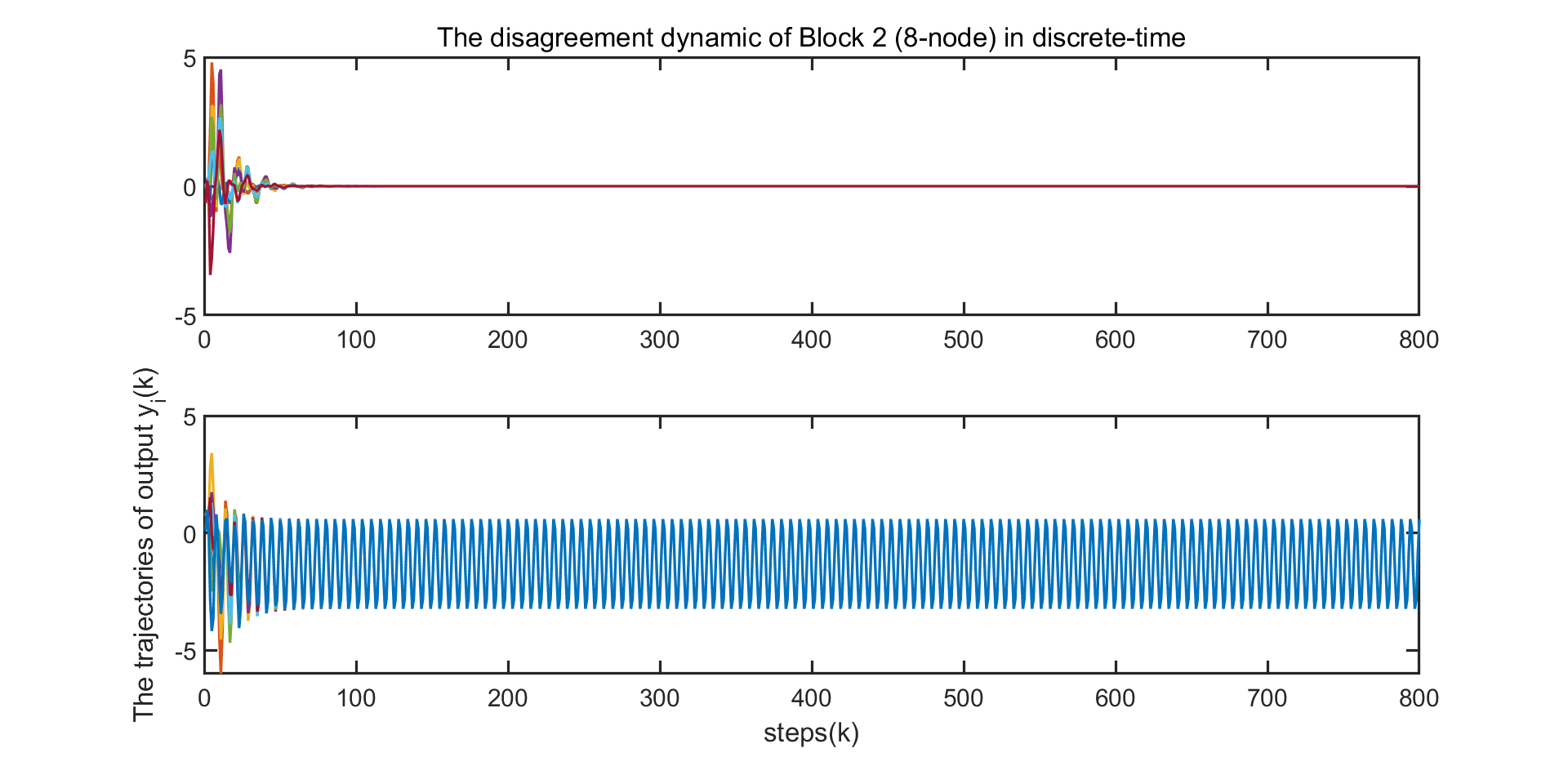} \centering
	\caption{Basic bicomponent 2 for discrete-time MAS: disagreement
		dynamic among the agents and synchronized output
		trajectories.}\label{dt8co} 
\end{figure}
\begin{figure}[ht]
	\includegraphics[width=9cm]{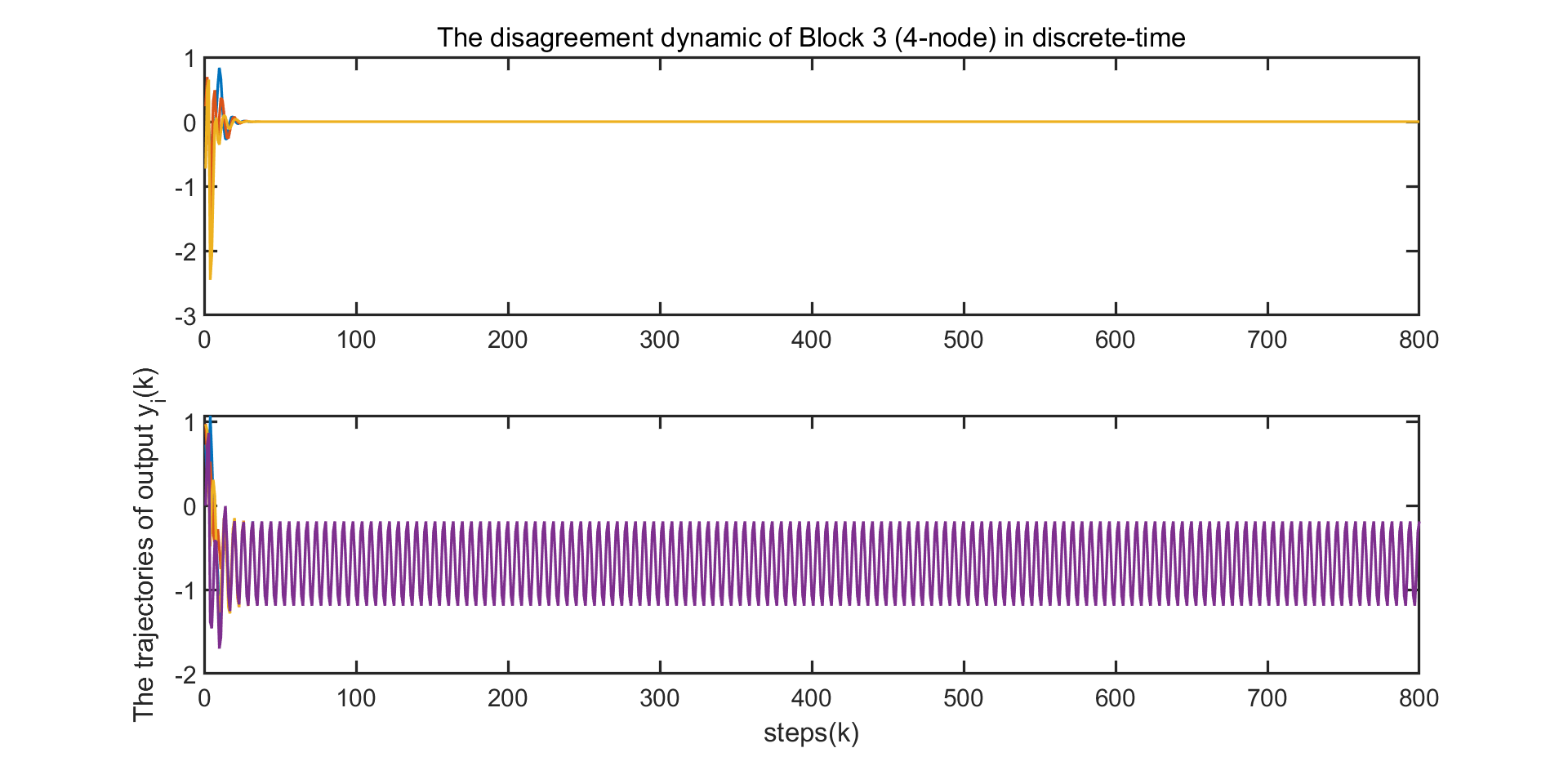}
	\centering
	\caption{Basic bicomponent 3 for discrete-time MAS: disagreement
		dynamic among the agents and synchronized output
		trajectories.}\label{dt1colead} 
\end{figure}

By using the scale-free protocol, we obtain $\zeta_i\to 0$ as
$t\to \infty$ see Figure \ref{diszeta60n}, which means weak
synchronization is achieved in the absence of connectivity. It implies
that the available network data for each agent goes to zero and
communication network becomes inactive.

We see that, consistent with the theory, we get output synchronization
within the three basic bicomponents as illustrated in figures
\ref{dt30co}, \ref{dt8co} and \ref{dt1colead}. Clearly,
the disagreement dynamic among the agents (the errors between the
output of agents) goes to zero within each basic
bicomponent. Similarly, we obtain that any agent outside of the basic
bicomponents converge to a convex combination of the synchronized
trajectories of the basic bicomponents 1, 2 and 3 based on Part 2
of Theorem \ref{thm_dy}.

\section{Conclusion}

In this paper we have introduced the concept of weak synchronization
for MAS.  We have shown that this is the right concept if you have no
information available about the network. If we have a directed
spanning tree it is equal to the classical concept of output
synchronization. However when, due to a fault, the network no longer
contains a directed spanning tree then we still achieve the best
synchronization properties possible for the given network. We have seen that the protocols
guarantee a stable response to these faults: within basic bicomponents
we still achieve synchronization and the outputs of the agents not
contained in a basic bicomponent converge to a
convex combination of the asymptotic behavior achieved in the basic
bicomponents. This behavior is completely independent of the specific
scale-free protocols being used.

For the heterogeneous MAS that we consider in this paper, the main
focus improving the available protocol design methodologies since for
heterogeneous networks the current designs are still limited in scope
but the concept of weak synchronization introduced in this paper is
the correct concept for this protocol design.

\bibliographystyle{plain}
\bibliography{referenc}

\end{document}